\documentclass[twocolumn,superscriptaddress,nobalancelastpage,10pt,pra,a4paper]{revtex4}
\usepackage{amsmath}
\usepackage{amsthm}
\usepackage{amssymb}
\usepackage{amsfonts}
\usepackage{graphicx}
\usepackage{wasysym}
\usepackage{mathrsfs}
\usepackage{yfonts}
\usepackage{bbold}
\usepackage{verbatim}
\usepackage{subfigure}
\usepackage{color}
\newcommand{\ket}[1]{\left| #1 \right\rangle}
\newcommand{\bra}[1]{\left\langle #1 \right|}
\newcommand{\scp}[2]{\langle #1 | #2 \rangle}

\newcommand{\opa}{\widehat{a}}

\newcommand{\opEps}{\widehat{E}^{(+)}}
\newcommand{\opEns}{\widehat{E}^{(-)}}

\newcommand{\x}{\mathbf{r},t}

\newcommand{\G}{\overleftrightarrow{G}^{(1)}}

\makeatletter
\newcommand*{\rom}[1]{\expandafter\@slowromancap\romannumeral #1@}
\makeatother

\begin{document}

%\begin{flushleft} \hfill \today \end{flushleft}

%\begin{center}
%\LARGE{\textbf{Controlling Partial Polarization of a Light Beam
%Quantum Mechanically}}
%\end{center}

%\begin{flushleft} \hfill{} \today
%\end{flushleft}

\title{Partial Polarization by Quantum Distinguishability}

\author{Mayukh Lahiri}
\email{mayukh.lahiri@univie.ac.at} \affiliation{Vienna Center for
Quantum Science and Technology (VCQ), Faculty of Physics,
Boltzmanngasse 5, University of Vienna, Vienna A-1090, Austria.}

\author{Armin Hochrainer} \affiliation{Vienna Center for
Quantum Science and Technology (VCQ), Faculty of Physics,
Boltzmanngasse 5, University of Vienna, Vienna A-1090, Austria.}

\author{Radek Lapkiewicz} \affiliation{Vienna Center for
Quantum Science and Technology (VCQ), Faculty of Physics,
Boltzmanngasse 5, University of Vienna, Vienna A-1090, Austria.}

\author{Gabriela Barreto Lemos} \affiliation{Vienna Center for
Quantum Science and Technology (VCQ), Faculty of Physics,
Boltzmanngasse 5, University of Vienna, Vienna A-1090,
Austria.}\affiliation{Institute for Quantum Optics and Quantum
Information, Austrian Academy of Sciences, Boltzmanngasse 3, Vienna
A-1090, Austria.}

\author{Anton Zeilinger}\affiliation{Vienna Center for Quantum Science and
Technology (VCQ), Faculty of Physics, Boltzmanngasse 5, University
of Vienna, Vienna A-1090, Austria.}\affiliation{Institute for
Quantum Optics and Quantum Information, Austrian Academy of
Sciences, Boltzmanngasse 3, Vienna A-1090, Austria.}

\begin{abstract}
Partial polarization is the manifestation of the correlation between
two mutually orthogonal transverse field components associated with
a light beam. We show both theoretically and experimentally that the
origin of this correlation can be purely quantum mechanical. We
perform a two-path first-order (single photon) interference
experiment and demonstrate that the degree of polarization of the
light emerging from the output of the interferometer depends on path
distinguishability. We use two independent methods to control the
distinguishability of the photon paths. While the distinguishability
introduced in one of the methods can be erased by performing a
suitable measurement on the superposed beam, the distinguishability
introduced in the other method cannot be erased. We show that the
beam is partially polarized only when both types of
distinguishability exist. Our main result is the dependence of the
degree of polarization on the inerasable distinguishability, which
cannot be explained by the classical (non-quantum) theory of light.

%\\ \\ PACS number(s):
\end{abstract}

\maketitle

%\vskip 2cm

%\tableofcontents

%\newpage
%\section{Aim}\label{sec:intro}
\noindent The interference effect displayed by a quantum system or
entity when sent through an interferometer is a key feature of
quantum mechanics \cite{Bohr-Einstein,Feynman-lec-3}. If quantum
entities of a particular kind are sent one at a time through a
two-path interferometer and one can identify (even in principle) the
path traversed by each of them, no interference occurs. The
information that leads to the identification of the path is often
called the path information. Because the common-sense understanding
of a particle implies that the path traversed can always be
identified, the particle behavior of the quantum entity is often
interpreted as the complete availability of the path information,
i.e, the complete distinguishability of the paths. On the other
hand, when the path information is fully unavailable, i.e, when the
paths are fully indistinguishable, perfect interference
occurs\textemdash a characteristic of waves.
\par
The wave-particle duality of a light quantum (photon) has been
confirmed by numerous experiments \cite{ZWJA-ph-interf-rev}.
Furthermore, the relationship between the path distinguishability
and the ability of light (or any quantum entity) to interfere has
been studied in detail \cite{GY-neutron-interf,M-ind,JSV,Eng-WPI}.
The ability of light to interfere has also long been the subject of
investigation in coherence theory, where it is quantified by the
degree of coherence \cite{Z-doc,W-coh,G-1,BW,MW}. It has been shown
that under certain circumstances, it is possible to establish a
relationship between the path distinguishability and the degree of
coherence \cite{M-ind}. The concepts of coherence theory are often
used to analyze polarization properties of light beams
\cite{W-pol,W-uni,LW-quant-pol,Note-Stokes-par}. It is therefore
natural to ask whether a connection between partial polarization and
path distinguishability can be established \cite{L-qind}.
\par
We develop a two-path interferometer in which we use two independent
methods for introducing path distinguishability in a controlled
manner. We investigate both theoretically and experimentally how the
polarization property of the light beam emerging from the output of
the interferometer depends on the distinguishability of photon
paths.
\par
\begin{figure}[htbp]  \centering
  \includegraphics[scale=0.4]{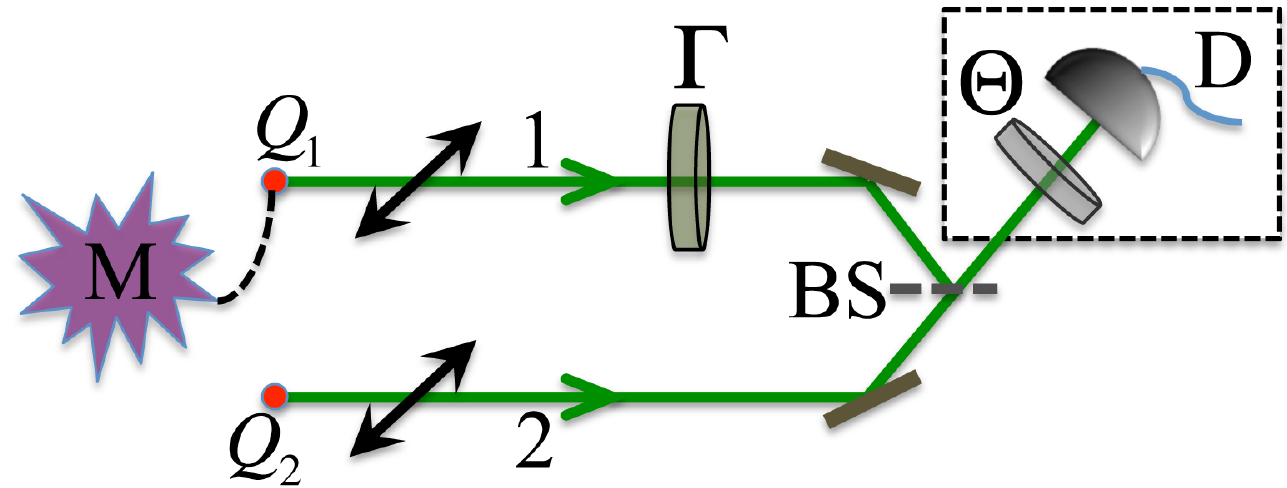}
  \qquad
\caption{Two identical sources, $Q_1$ and $Q_2$, emit
indistinguishable linearly polarized photons. The photons generated
by $Q_1$ are sent through a polarization rotator, $\Gamma$, and then
superposed with the photons generated by $Q_2$ by a beam splitter,
BS. The superposed beam emerging from one of the outputs of BS is
sent through a polarizer, $\Theta$, and then detected by a
photo-detector, D. The device, M, attached to $Q_1$ determines with
probability $1-\mathcal{M}^2$ whether $Q_1$ has emitted.}
\label{fig:two-source-interf-par-pol-ed}
\end{figure}

Let us begin by considering a thought experiment in which photon
beams generated by two identical sources $Q_1$ and $Q_2$ are
superposed by a lossless and balanced non-polarizing beam-splitter,
BS (Fig. \ref{fig:two-source-interf-par-pol-ed}).
Photons emerging from one of the outputs of BS are collected by a
photo-detector, D. Clearly, this system is equivalent to a two-path
interferometer. The two paths are of equal length and we label them
as 1 and 2. Photons emitted by $Q_1$ can only travel via path 1 and
photons emitted by $Q_2$ can only travel via path 2. We assume that
the photons emitted by $Q_1$ and $Q_2$ are linearly polarized in the
same direction and are identical to each other. On path 1, we place
a polarization rotator (for example, a half-waveplate), $\Gamma$, by
which we can rotate the direction of the incident linear
polarization by an arbitrary angle $\gamma$, where $\cos\gamma\geq
0$ \cite{Note-gamma}; the superposed beam emerging from BS is sent
through a polarizer, $\Theta$, before arriving at D [Fig.
\ref{fig:two-source-interf-par-pol-ed}]. The light emerging from
$\Theta$ is linearly polarized along a direction that makes an angle
$\theta$ with the polarization-direction of the light originally
emitted by the sources. In this arrangement, $\Theta$ and D
constitute the detection system that measures the degree of
polarization of the superposed beam emerging from BS.
\par
Suppose $Q_1$ and $Q_2$ emit at the same rate and in such a way that
only one photon exists in the system between an emission and a
detection at D. Now, if path 2 is blocked, the probability amplitude
of photo-detection is given by
$\alpha_1=e^{i\phi_1}[\cos(\theta-\gamma)]/2$, where the phase
$\phi_1$ depends on path length. Similarly, if path 1 is blocked,
the probability amplitude of photo-detection will be
$\alpha_2=e^{i\phi_2}(\cos\theta)/2$. Suppose now that we attach
(Fig. \ref{fig:two-source-interf-par-pol-ed}) to source $Q_1$ a
device, M, that does \emph{not} perform any measurement on the
photons entering the interferometer but determines with a known
probability whether $Q_1$ has emitted. Clearly, when both paths are
open, there are three possible cases in which a photon can arrive at
D: \rom{1}) $Q_1$ emits and M reports the emission; \rom{2}) $Q_1$
emits and M does not report the emission; \rom{3}) $Q_2$ emits.
Since the photons emitted by $Q_1$ and $Q_2$ are identical,
possibilities \rom{2} and \rom{3} are indistinguishable. On the
other hand, possibility \rom{1} is fully distinguishable from
possibilities \rom{2} and \rom{3}. For obtaining the total
probability of detecting a photon at D, one therefore needs to add
the probability associated with possibility \rom{1} to the modulus
square of the sum of \emph{probability amplitudes} associated with
\rom{2} and \rom{3}. Let us assume that when $Q_1$ has emitted a
photon, the probability of M \emph{not} reporting the emission is
equal to $\mathcal{M}^2$, where $0\leq \mathcal{M}\leq 1$. The
probability amplitudes associated with cases \rom{1}, \rom{2}, and
\rom{3} are then given by $\alpha_1 \sqrt{1-\mathcal{M}^2}$,
$\alpha_1 \mathcal{M}$, and $\alpha_2$, respectively. The
probability of photo-detection at D is thus given by
\begin{align}\label{prob-ph-det-indist-pol}
&\Phi= |\alpha_1
\sqrt{1-\mathcal{M}^2}|^2+|\alpha_1\mathcal{M}+\alpha_2|^2 \nonumber \\
&=\frac{1}{4}\big[\cos^2\theta+\cos^2(\theta-\gamma) \nonumber \\
& \qquad  \qquad +2\mathcal{M}
\cos\theta\cos(\theta-\gamma)\cos(\phi_2-\phi_1)\big],
\end{align}
which is directly proportional to the photon counting rate at the
detector.
\par
It is to be noted that the path distinguishability has been
introduced by two independent methods: a) by the polarization
rotator $\Gamma$, and b) by the device M. The path
distinguishability introduced by $\Gamma$ can be erased by the
polarizer $\Theta$, as is evident from the term
$\cos(\theta-\gamma)$ of Eq. (\ref{prob-ph-det-indist-pol}). When
$\gamma=\pi/2$ and $\theta=0$, i.e., when the maximum path
distinguishability is introduced by $\Gamma$ and the
distinguishability is not erased by $\Theta$, no interference occurs
for any value of $\mathcal{M}$. It is clear that a measure of the
path distinguishability (or, equivalently, indistinguishability)
introduced by $\Gamma$ can be represented by $\cos\gamma$. On the
other hand, the path distinguishability introduced by M cannot be
erased by any other device. A measure of the path distinguishability
introduced M is given by $\mathcal{M}$. When $\mathcal{M}=0$, M
determines with complete certainty whether $Q_1$ has emitted. In
this case, the paths are fully distinguishable and no interference
occurs, irrespective of the orientation of $\Gamma$.
\par
For the sake of simplicity, we choose the path lengths such that
$\phi_2-\phi_1$ is equal to a multiple of $2\pi$, i.e.,
$\cos(\phi_2-\phi_1)=1$. In this case, it can be readily shown from
Eq. (\ref{prob-ph-det-indist-pol}) that when $\theta=\gamma/2$, the
probability $\Phi$ attains its maximum value
$\Phi_{\text{max}}=(1+\mathcal{M})\cos^2(\gamma/2)/2$; and when
$\theta=\gamma/2\pm \pi/2$, it attains the minimum value
$\Phi_{\text{min}}=(1-\mathcal{M})\sin^2(\gamma/2)/2$. The degree of
polarization of the beam generated by superposition is given by
\cite{Note-dop-vis}
\begin{align}\label{dop-interm}
P=\frac{\Phi_{\text{max}}-\Phi_{\text{min}}}{\Phi_{\text{max}}
+\Phi_{\text{min}}}=
\frac{\mathcal{M}+\cos\gamma}{1+\mathcal{M}\cos\gamma}.
\end{align}
It follows from Eq. (\ref{dop-interm}) that \emph{both} devices
($\Gamma$ and M) must be used to introduce path distinguishability
for generating a partially polarized ($0<P<1$) beam. If only one of
the devices introduces path distinguishability, i.e., if either
$\mathcal{M}=1$ or $\cos\gamma=1$, the beam is always fully
polarized ($P=1$). On the other hand, the beam is unpolarized
($P=0$) if and only if \emph{both} devices introduce maximum path
distinguishability, i.e., if and only if $\mathcal{M}=0$ and
$\cos\gamma=0$. The central feature of the thought experiment is the
dependence of the degree of polarization on the inerasable
distinguishability for a ``fixed amount of the erasable
distinguishability,'' because the phenomenon is beyond the scope of
the classical theory of optical fields.
\par
\begin{figure}[htbp]  \centering
  \includegraphics[scale=0.4]{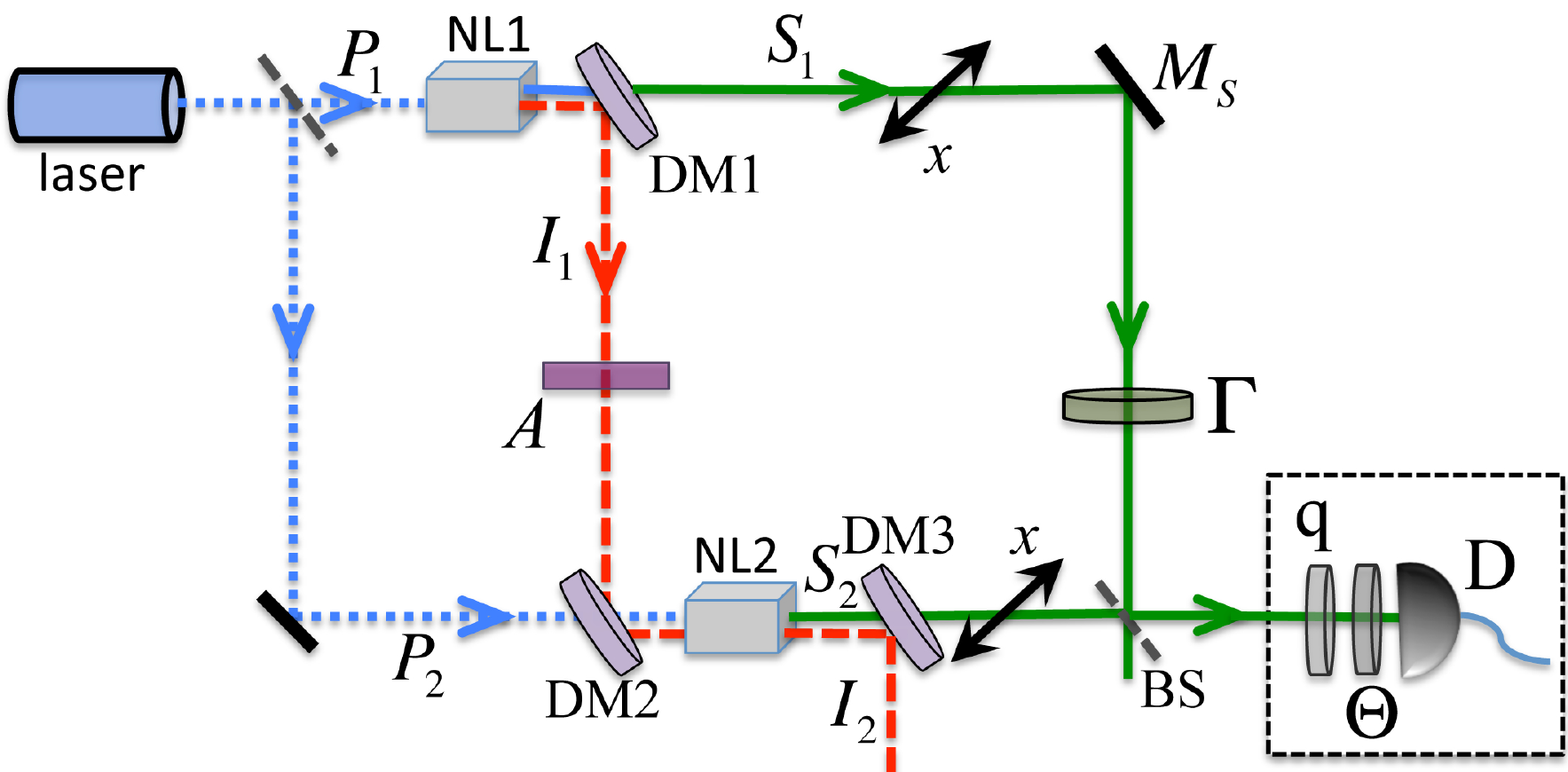}
  \qquad
\caption{Schematic diagram of the experimental setup. Pump beams
$P_1$ and $P_2$ (dotted lines) are generated by a CW laser source
(532 nm). Two identical nonlinear crystals, NL1 and NL2, produce
linearly polarized signal ($S_1$, $S_2$; solid line) and idler
($I_1$, $I_2$; dashed line) photons with mean wavelengths 810 nm and
1550 nm, respectively. A neutral density filter, $A$, is placed on
the path of $I_1$ between NL1 and NL2. Initially, $S_1$ and $S_2$
are polarized in the same direction $x$. Half-wave plate $\Gamma$
rotates the polarization direction of $S_1$ by an angle $\gamma$.
$S_1$ and $S_2$ are superposed by a non-polarizing beam splitter,
BS. The detection-system (inside the dotted box) used for
polarization state tomography consists of a quarter-wave plate, q, a
polarizer, $\Theta$, and an avalanche photo detector, D. DM1, DM2
and DM3 are dichroic mirrors used to separate/combine signal, idler
and pump beams.} \label{fig:set-up-pol-schm-ed}
\end{figure}

We now discuss an experiment in which the above-mentioned phenomenon
is observed. This experiment is based on the concept of so-called
``induced coherence without induced emission''
\cite{ZWM-ind-coh-PRL,WZM-ind-coh-PRA}. Two identical nonlinear
crystals, NL1 and NL2, are pumped by two mutually coherent pump
beams, $P_1$ and $P_2$, respectively (Fig.
\ref{fig:set-up-pol-schm-ed}). Each crystal converts a pump photon
into a photon pair (signal and idler), each \emph{linearly
polarized}, by the process of parametric down-conversion. We denote
the signal and the
idler generated in NL1 by $S_1$ and $I_1$, respectively, and the
signal and the idler generated in NL2 by $S_2$ and $I_2$,
respectively. The idler beam, $I_1$, is sent through NL2 and is
aligned with $I_2$ (Fig. \ref{fig:set-up-pol-schm-ed}). We ensure
that the down-converted light is weak enough, so that it is highly
improbable for photon pairs emitted by both crystals to be
simultaneously present in the system. Under this condition the
effect of stimulated emission at NL2 can be neglected. An attenuator
(neutral density filter), $A$, is placed on the path of $I_1$
between NL1 and NL2; the transmission coefficient of $A$ can be
varied. The signal beam $S_1$ is sent through a half-wave plate,
$\Gamma$, such that its polarization direction can be rotated by a
chosen angle $\gamma$. It is then superposed with $S_2$ by a
non-polarizing beam-splitter, BS.
\par
We first need to meet the condition under which the beams $S_1$ and
$S_2$ interfere (i.e., the beams are mutually coherent) in absence
of $A$ and $\Gamma$. According to the principles of quantum
mechanics, if there exists any information leading to the
identification of the path traversed by a signal photon emerging
from BS, no interference occurs. Although the signal photons emitted
by the two crystals are identical in all aspects (i.e., same
polarization, frequency, etc.), it is possible to obtain the path
information by the method of coincidence detection, simply because
signal and idler of a down-converted photon pair are produced
``simultaneously'' at a particular crystal. However, it has been
shown that under a certain condition this path information can be
removed \cite{WZM-ind-coh-PRA}. Suppose that $\tau_{S_1}$,
$\tau_{S_2}$, and $\tau_{I_1}$ are the propagation times of $S_1$
from NL1 to D, of $S_2$ from NL2 to D, and of $I_1$ from NL1 to NL2.
If $|\tau_{S_1}-\tau_{S_2}-\tau_{I_1}|$ is less than the coherence
time of the down converted light, it is not possible to distinguish
between an $S_1$ photon and an $S_2$ photon at D in absence of $A$
and $\Gamma$. We set the optical path lengths in our experiment in
such a way that this condition is met. We stress that measurement of
coincidence counts is absolutely not required in order to observe
the interference at $D$, because the unavailability of path
information is alone enough for this purpose. In fact, \emph{no
coincidence measurement is performed in our experiment}. It must
further be noted that $S_1$ and $S_2$ beams interfere \emph{not}
because of stimulated emission occurring at NL2
\cite{ZWM-ind-coh-PRL,WZM-ind-coh-PRA,WM-ind-coh-cl-q}.
\par
Once the appropriate optical path lengths are chosen, the attenuator
($A$) and the half-wave plate ($\Gamma$) are placed at their
positions (Fig. \ref{fig:set-up-pol-schm-ed}). The degree of
polarization of the superposed signal beam emerging from one of the
outputs of BS is determined. Below we provide both the theoretical
analysis and the experimental results.
\par
The photons $S_1$ and $S_2$ are generated as linearly polarized in
the same direction $x$, say; $I_1$ and $I_2$ are linearly polarized
along the direction $x'$, say \cite{Note-pol-dirs}. The quantum
state of light (interaction picture \cite{MW}) generated by a
crystal is given by the well-known formula
\begin{equation}\label{state-pdc-int-series-pol-sm}
[\mathbb{1}+(g_j\opa_{S_jx}^{\dag}
\opa_{I_jx'}^{\dag}-\text{H.c.})+\dots]\ket{\psi_{j0}}\equiv
\widehat{U}_j\ket{\psi_{j0}},
\end{equation}
where $\mathbb{1}$ is the identity operator, $j=1,2$ labels the
crystals, $g_j$ provides a measure of the rate of parametric
down-conversion, $\opa_{S_jx}^{\dag}$ and $\opa_{I_jx'}^{\dag}$ are
creation operators for $S_j$ and $I_j$ photons, respectively, H.c.
represents Hermitian conjugation, and $\ket{\psi_{j0}}$ is the state
of light before down-conversion (input state). The action of the
attenuator, $A$, on the quantized field associated with
$I_1$-photons is equivalent to that of a lossless beam splitter
\cite{ZWM-ind-coh-PRL,WZM-ind-coh-PRA}. One thus obtains the
following relation:
\begin{align}\label{a-idl-op-rel-sp-sm-pol}
\opa_{I_2x'}=\left[T \opa_{I_1x'} +R'\opa_{0x'} \right]~e^{i\phi_I},
\end{align}
where $T$ is the complex amplitude transmission coefficient of $A$,
$|T|^2+|R'|^2=1$, $\opa_{0x'}$ represents the vacuum field at the
unused port of the beam splitter (the attenuator $A$), and $\phi_I$
is a phase factor due to propagation of $I_1$ from NL1 to NL2. It
follows from Eqs. (\ref{state-pdc-int-series-pol-sm}) and
(\ref{a-idl-op-rel-sp-sm-pol}) that the quantum state of light
generated in this system is given by
\begin{align}\label{q-state-system}
\ket{\Psi}=&\widehat{U}_2\widehat{U}_1 \ket{\text{vac}}\nonumber \\
\approx &\ket{\text{vac}}+\left(g_1\ket{x}_{S_1} +g_2
e^{-i\phi_I}T^{\ast}\ket{x}_{S_2}\right)\ket{x'}_{I_1} \nonumber \\
&+g_2 e^{-i\phi_I}R'^{\ast}\ket{x}_{S_2}\ket{x'}_{0},
\end{align}
where $\ket{\text{vac}}$ is the vacuum state, $\ket{x}_{S_j}\equiv
\opa_{S_jx}^{\dag}\ket{\text{vac}}$ represents an $x$-polarized
signal photon, $\ket{x'}_{I_j}\equiv
\opa_{I_jx'}^{\dag}\ket{\text{vac}}$ represents an $x'$-polarized
idler photon, $\ket{x'}_{0}\equiv
\opa_{0x'}^{\dag}\ket{\text{vac}}$, $_0\scp{x'}{x'}_0=1$, and we
have neglected the higher order terms.
\par
The positive-frequency part of the quantized field components,
associated with the superposed beam emerging from one of the outputs
of the beam-splitter, can be represented by
\begin{subequations}\label{field-D-th}
\begin{align}
\opEps_{Sx} &=e^{i\phi_{S1}}\left(\cos\gamma~\opa_{S_1x} -\sin\gamma
~\opa_{S_1y} \right) +ie^{i\phi_{S2}}~\opa_{S_2x},
\label{field-D-th:a} \\
\opEps_{Sy} &=e^{i\phi_{S1}}\left(\sin\gamma ~\opa_{S_1x}+
\cos\gamma ~\opa_{S_1y} \right) +ie^{i\phi_{S2}}~\opa_{S_2y},
\end{align}
\end{subequations}
where $y$ is the Cartesian direction orthogonal to $x$, $\phi_{S1}$
and $\phi_{S2}$ are the phase changes associated with the
propagation from NL1 to BS and from NL2 to BS, respectively, and we
have included the action of $\Gamma$ on $S_1$. The degree of
polarization can be determined by using the formula
\cite{LW-quant-pol,Note-dop-def}
\begin{equation}\label{dop-def}
P=\{1-4~\text{det}~~ \G/[\text{tr}~\G]^2\}^{1/2},
\end{equation}
where det and tr represent the determinant and the trace of a
matrix, respectively, and $\G$ is a $2\times 2$ correlation matrix
whose elements are given by \cite{G-1,LW-quant-pol,Note-coh-mat}
\begin{equation}\label{G-pol-def}
G^{(1)}_{pq}=\bra{\Psi}\opEns_{sp}\opEps_{sq}\ket{\Psi}.
\end{equation}
Here $p=x,y$, $q=x,y$, the quantum state $\ket{\Psi}$ is given by
Eq. (\ref{q-state-system}), the quantized field components
$\opEps_{p}(\x)$ are given by Eqs. (\ref{field-D-th}), and
$\opEns_{p}(\x)= \{\opEps_{p}(\x)\}^{\dag}$.
\par
If $|g_1|=|g_2|$, i.e., if the two crystals emit at the same rate,
one can show by the use of Eqs. (\ref{q-state-system})\textendash
(\ref{G-pol-def}) that the degree of polarization of the superposed
signal beam is given by
\begin{align}\label{dop-form-gen}
P=\big\{& \cos^2\gamma+|T|^2(\sin^2\gamma +\cos^2\gamma
\cos^2\beta) \nonumber \\
& +2|T|\cos\gamma \cos\beta \big\}^{\frac{1}{2}}
/\big\{1+|T|\cos\gamma \cos\beta\big\},
\end{align}
where
$\beta=\phi_{S2}-\phi_{S1}-\phi_I-\text{arg}(T)+\text{arg}(g_2)
-\text{arg}(g_1)$. If we set $\cos\beta=1$, Eq. (\ref{dop-form-gen})
reduces to the form
\begin{equation}\label{dop-form-sp}
P=\frac{|T|+\cos\gamma}{1+|T|\cos\gamma}.
\end{equation}
\par
It is to be noted that Eq. (\ref{dop-form-sp}) is strikingly similar
to Eq. (\ref{dop-interm}). This is due to the following reasons:
When $I_1$ beam passes through $A$, its intensity drops by a factor
of $|T|^2$. Since a photon cannot be broken into further fractions,
an idler photon can either be fully transmitted or fully blocked by
$A$. The probability of an idler photon being transmitted through
$A$ is therefore equal to $|T|^2$. If the idler photon is
transmitted, one looses the path information for the signal photon
completely (in absence of $\Gamma$); and if it is blocked, the path
information for the signal photon is fully available. Because of
this, the attenuator, $A$, of Fig. \ref{fig:set-up-pol-schm-ed}
plays the role of the device, M, shown in Fig.
\ref{fig:two-source-interf-par-pol-ed}. The half-wave plate,
$\Gamma$, plays the same role in both cases. Therefore, if we set
the interferometric phase to a multiple of $2\pi$, the degree of
polarization of the superposed signal beam is obtained from Eq.
(\ref{dop-interm}) just by replacing $\mathcal{M}$ with $|T|$.
\begin{figure} \centering
 \subfigure[] {
    \label{fig:dop-T-plot-exp}
    \includegraphics[scale=0.25]{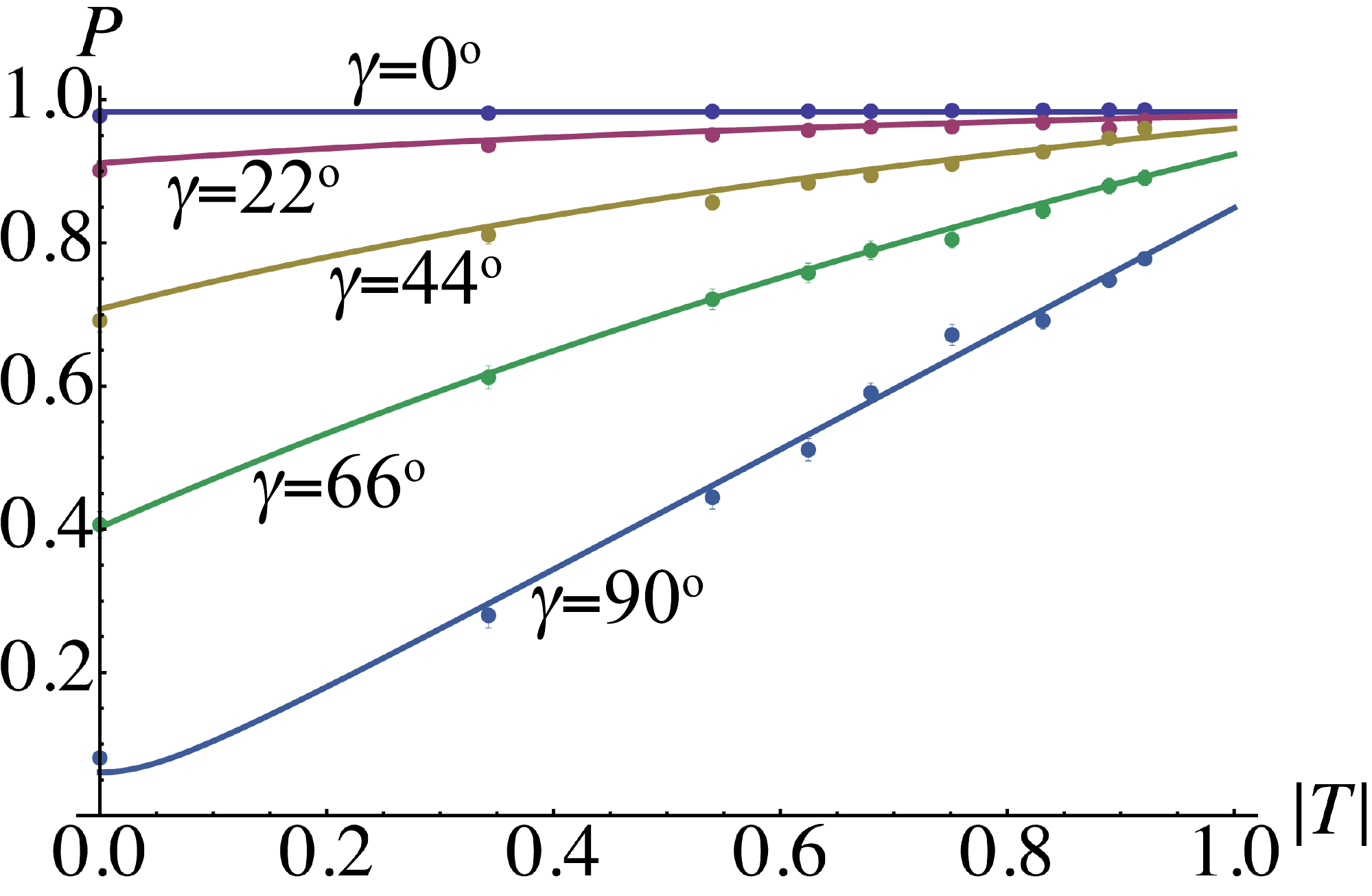}
} \hskip 0.5cm
   \subfigure[] {
    \label{fig:dop-T-plot-th}
    \includegraphics[scale=0.225]{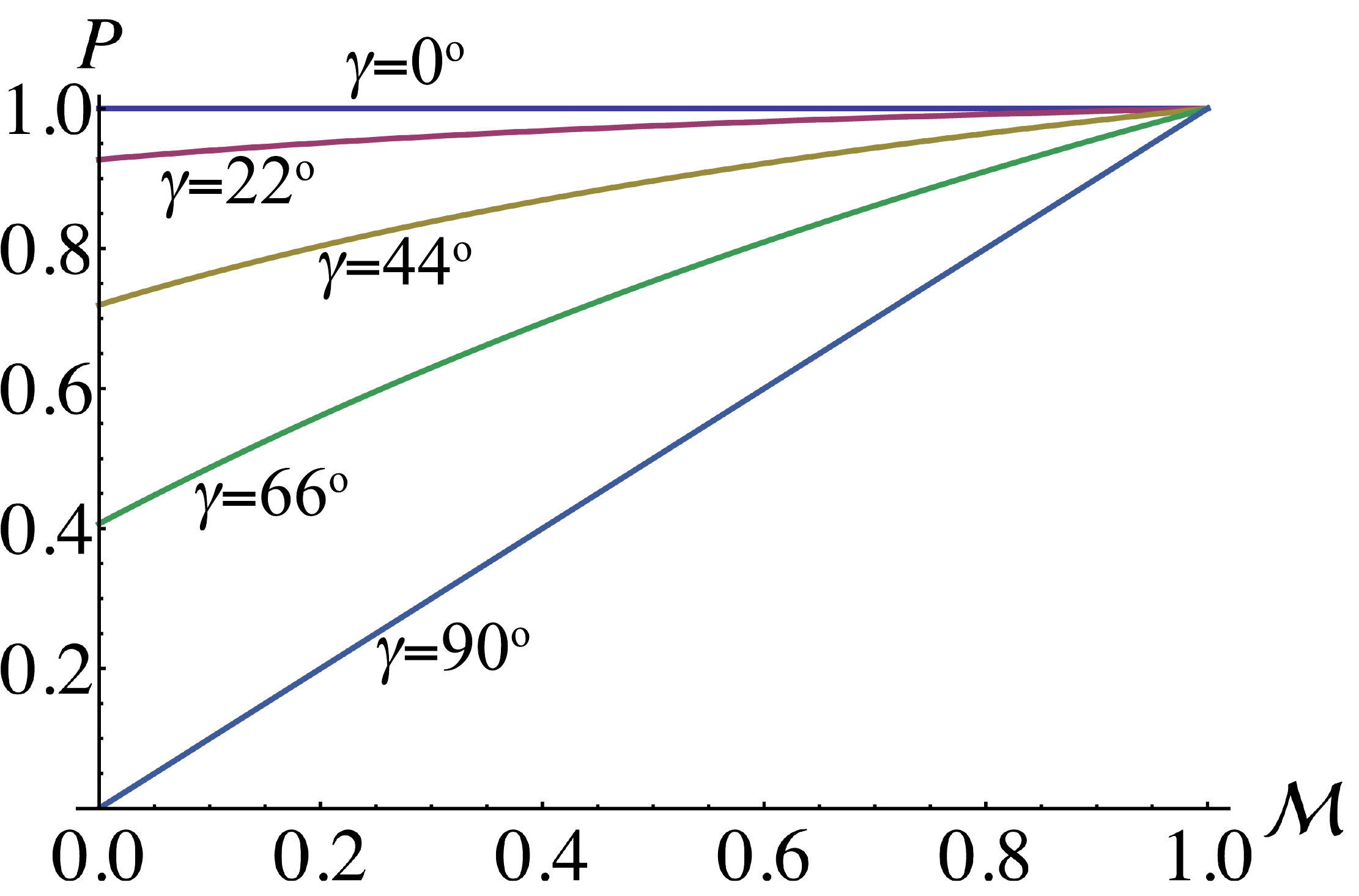}
} \caption{Dependence of the degree of polarization $P$ on
inerasable path distinguishability: (a) Experimentally observed
dependence of $P$ on $|T|$ for various values of $\gamma$, and
computed curves considering experimental imperfections (solid
lines). Data points are represented by filled circles with error
bars including both systematic and statistical errors. (b)
Dependence of $P$ on $\mathcal{M}$ for various values of $\gamma$
predicted by the thought experiment [Eq. (\ref{dop-interm})].}
\label{fig:dop-T-plot}
\end{figure}
\par
In the experiment (Fig. \ref{fig:set-up-pol-schm-ed}), the value of
$|T|$ is changed by inserting neutral density filters ($A$) of
different values of transmittance in $I_1$ beam. For each choice of
$T$ and $\gamma$, the interferometric phase is set equal to a
multiple of $2\pi$ by varying the path-length of $S_1$ with the
mirror $M_S$ placed on a piezo-driven stage. Polarization state
tomography is performed on the superposed signal beam by using a
quarter waveplate, q, a polarizer, $\Theta$, and a single-photon
counting module, D. The photon counting rate at D is recorded for 15
secs in each measurement. The data are corrected for background
(dark counts), and the matrix $\G$ is determined by the maximum
likelihood technique for a single-qubit system
\cite{JKMW-qbit-tom-PRA,AJK-qstate-tom-chap,Note-ml-tech}. The
degree of polarization is then calculated by using Eq.
(\ref{dop-def}).
\par
The experimental results are shown in Fig. \ref{fig:dop-T-plot-exp}.
When $\gamma=90^{\circ}$ and $|T|=0$, the experimentally measured
degree of polarization is slightly more than zero; this is due to
the fact that the non-polarizing beam splitter (BS) is in practice
slightly polarizing. Also for $\gamma\neq 0$ and $|T|=0$, we do not
obtain completely polarized light; this is because under the
experimental conditions, the $I_1$ beam suffers losses at the
various optical components through which it passes. The solid curves
of Fig. \ref{fig:dop-T-plot-exp} are obtained by computation
considering all these experimental imperfections. A comparison of
Figs. \ref{fig:dop-T-plot-exp} and \ref{fig:dop-T-plot-th} shows
that the predictions of the thought experiment (Fig.
\ref{fig:two-source-interf-par-pol-ed}) have been practically
realized in the actual experiment (Fig.
\ref{fig:set-up-pol-schm-ed}).
\par
It is important to understand that the attenuator, $A$, introduces
path distinguishability for the signal photons \emph{without
interacting with them}. This path distinguishability (quantified by
$|T|$) cannot be erased by introducing any device that interacts
with the signal photons (in this context, see \cite{KSC-q-eraser}).
This is what renders this experiment beyond the scope of the
classical theory of optical fields. In the special case of
$\gamma=\pi/2$, the degree of polarization is equal to $|T|$ (see
the curves labeled by $\gamma=90^{\circ}$ in Fig.
\ref{fig:dop-T-plot}). This value of the degree of polarization
cannot be increased by, for example, frequency filtering the
superposed beam, i.e., by enhancing the coherence time of the light.
\par
In the classical theory, light is considered as an electromagnetic
wave and its properties are described by the fluctuating
electromagnetic field associated with it. Partial polarization of a
light beam is considered to be a manifestation of correlation
between the transverse field components (\cite{BW}, Sec. 10.9). The
classical theory of partial polarization is based on the assumption
that stochastic fluctuations are always associated with optical
fields. We have demonstrated that partial polarization of a light
beam can be solely due to the wave-particle duality of photons,
i.e., solely due to the quantum nature of light. However, many
simplest-order correlation properties of the partially polarized
beam generated in our experiment can be successfully explained by
the use of the classical theory of statistical optics. It is thus
clear that even if several properties of the beam can be explained
by assuming the existence of correlations between stochastic
classical fields, the origin of such correlations may not always be
explained by the classical theory of radiation, but only by quantum
mechanics.
\par
The research was supported by \"OAW, the European Research Council
(ERC Advanced grant no. 227844 ``QIT4QAD'', and SIQS grant no.
600645 EU-FP7-ICT), and the Austrian Science Fund (FWF) with SFB F40
(FOQUS) and W1210-2 (CoQus).

% (The chosen values of $|T|$ are:
%$|T|\approx 0,0.34,0.54,0.62,0.68,0.75,0.83,0.89,0.92$. These values
%of are measured by using a laser beam whose mean wavelength (1550
%nm) is equal to that of the idler beam and a power meter.)

%\newpage


\begin{thebibliography}{99}
\bibitem {Note-Stokes-par} Alternatively, partial polarization is also
studied by the use of the Stokes parameters
\cite{Stokes-param-orig,Fano-pol}. Although Stokes did not use any
statistical theory to introduce these parameters, the classical
Stokes parameters can be represented in terms of correlation
functions of optical field components \cite{BW,MW}.
\bibitem {Note-gamma} In order to cover all possible directions of linear polarization, it
is sufficient to assume that $\gamma$ lies in the range $-\pi/2\leq
\gamma \leq \pi/2$.
\bibitem {Note-dop-vis} This formula is obtained from Eq. (\ref{dop-def}) by analogy with a
standard method used in the classical theory of partial polarization
[see \cite{BW}, Sec. 10.9, Eqs. (20) and (52)]. It can be shown that
such an analogy is justified \cite{LW-quant-pol}.
\bibitem {Note-pol-dirs} Although in our experiment the directions $x$ and $x'$
are the same, in the theoretical analysis, we choose them to be
different for the sake of generality.
\bibitem {Note-dop-def} This formula is the quantum mechanical version of the standard
formula of the degree of polarization used for classical fields
(\cite{BW}, Sec. 10.9.2, Eq. (52)). If the state of polarization is
represented by a point in the Poincar\'e sphere, $P$ represents the
distance of the point from the center of the sphere.
\bibitem {Note-coh-mat} This matrix is equivalent to Wiener's coherency
matrix \cite{NW-coh-mat}.
\bibitem {Note-ml-tech} This method actually determines the normalized correlation matrix,
which is practically equivalent to the original correlation matrix
used for determining the degree of polarization.
\end{thebibliography}
\end{document}